\documentclass{article}
\usepackage{spconf,amsmath,graphicx}
\usepackage{subfigure}
\usepackage{color}
\usepackage{cleveref}

\usepackage{mathtools}
\usepackage{amssymb}
\usepackage{tabularx}

\title{Sparse Gaussian process Audio Source Separation Using Spectrum Priors in the Time-Domain}
\name{Pablo A. Alvarado$^{\star}$\sthanks{Supported by Colciencias scholarship 679.} \qquad Mauricio A. \'{A}lvarez$^{\mathsection}$ \qquad Dan Stowell$^{\star}$\sthanks{Supported by EPSRC Fellowship EP/L020505/1.} }
\address{$^{\star}$Centre for Digital Music, Queen Mary University of London, London, UK \\ $^{\mathsection}$Department of Computer Science, The University of Sheffield, Sheffield, UK}
\begin{document}
%
\maketitle
\begin{abstract}
Gaussian process (GP) audio source separation is a time-domain approach that circumvents the inherent phase approximation issue of spectrogram based methods. 
Furthermore, through its kernel, GPs elegantly incorporate prior knowledge about the sources into the separation model. 
Despite these compelling advantages, the computational complexity of GP inference scales cubically with the number of audio samples.
As a result, source separation GP models have been restricted to the analysis of short audio frames.
We introduce an efficient application of GPs to time-domain audio source separation, without compromising performance. 
For this purpose, we used GP regression, together with spectral mixture kernels, and variational sparse GPs.
%
%
We compared our method with LD-PSDTF (positive semi-definite tensor factorization),  KL-NMF (Kullback-Leibler non-negative matrix factorization), and IS-NMF (Itakura-Saito NMF).
Results show that the proposed method outperforms these techniques. 
\end{abstract}

\section{Introduction}
Single-channel audio source separation is a central problem in signal processing research. 
Here, the task is to estimate a certain number of latent signals or \textit{sources} that were mixed together in one recorded \textit{mixture} signal \cite{Liutkus11}. 
State of the art time-frequency methods for source separation include non-negative matrix factorisation (NMF) \cite{Lee00}, and probabilistic latent component analysis (PLCA) \cite{Smaragdis07}. 
These approaches decompose the power spectrogram of the mixture into elementary components. 
Then, the components are used to calculate the individual source-spectrograms. 
Time-frequency methods often arbitrarily discard phase information.
As a result, the phase of each source-spectrogram must be approximated, corrupting the reconstructed sources.

In contrast, time-domain source separation approaches can avoid the phase approximation issue of time-frequency methods \cite{Fevotte14, Stoller18}. 
For example, Yoshii et al. \cite{Yoshii13} reconstructed source signals  from the mixture waveform directly in the time domain. 
To this end, Gaussian processes (GPs) were used to predict each source waveform.
GPs are probability distributions over functions \cite{rasmussen05}.
A Gaussian process is completely defined by a mean function, and a kernel or covariance function.
In fact, the kernel determines the properties of the functions sampled from a GP.
A particularly influential work in time domain approaches is Liutkus et al. \cite{Liutkus11}, who first formulated source separation as a GP regression task. 
%

%
Although source separation Gaussian process (SSGP) models circumvent phase approximation,
the computational complexity of GP inference scales cubically with the number of audio samples.
%
Hence, different approximate techniques have been proposed to make the separation tractable.
For instance, various authors partitioned the mixture signal into independent frames \cite{Liutkus11, Yoshii13}.
Further, approximate inference in the frequency domain was used to learn model hyperparameters \cite{Liutkus11}. 
%
%
Alternatively, Adam et al. \cite{Adam16} recently proposed to use variational sparse GPs for source separation, however audio signals were beyond the scope of their study.
Variational approaches rely on a set of \textit{inducing variables} to build a low-rank approximation of the full covariance matrix.
Here, the approximate distribution and hyperparameters are learned together by maximising a lower bound of the true marginal likelihood \cite{Titsias09}. 
Moreover, variational inference has allowed the application of GPs models to large datasets \cite{Hensman13, Hensman18}.

%
Despite the kernel selection in SSGP models determines 
the properties of sources, only standard covariance functions have been used so far.
For example, Adam et al. \cite{Adam16} considered stationarity, smoothness and periodicity, using \textit{exponentiated quadratic} times \textit{cosine} kernels. 
\textit{Standard periodic} kernels \cite{Mackay98}  were applied in \cite{Liutkus11}.
%
These kernels assume that the source spectrum is composed of a fundamental frequency and perfect harmonics. 
However, real audio signals have more intricate spectra \cite{Berg14},
and so separating audio sources requires more flexible covariance functions. 
One such covariance, the spectral mixture (SM) kernel \cite{Wilson13}, is intended for intricate spectrum patterns.
SM kernels approximate the spectral density of any stationary covariance function, using a Gaussian mixture. 
Alternatively, non-parametric kernels are implicitly considered when the covariance matrix of each source is directly optimised by maximum likelihood \cite{Yoshii13}.
However, that study did not contemplate variational sparse GPs. 
%
To our knowledge, it has not been determined whether incorporating SM kernels together with variational sparse GPs into source separation models leads to more efficient and accurate audio source reconstructions.

In this paper we introduce a method that combines GP regression \cite{rasmussen05, Liutkus11}, spectral mixture kernels \cite{Wilson13}, and variational sparse GPs \cite{Titsias09}. 
We consider the mixture data as noisy observations of a function of time, 
composed as the sum of a known number of sources. 
Further, we assume that each source follows a different GP with a distinctive spectral mixture kernel.
In addition, we adapt the kernels to reflect prior knowledge about the typical spectral content of each source. 
%
%
Also, we frame the mixture data, and for every frame we maximize a variational lower bound of the true marginal likelihood to learn the hyperparameters that control the amplitude of each source.
Finally, to separate the sources, we use the learned priors to calculate the true posterior over each source.
%

\section{Gaussian process source separation}

We notate the mixture data as $\mathbf{y} = \left[ y_i \right]_{i=1}^{n}$ at time instants $\mathbf{t} = \left[ t_i \right]_{i=1}^{n}$.
As mentioned previously, we consider each mixture audio sample $y_i$ as an observation of a mixture function $f(t)$ corrupted by independent Gaussian noise. 
Further, we assume $f(t)$ as the sum of $J$ independent source functions $\left\lbrace s_j(t) \right\rbrace_{j=1}^{J} $. 
These functions represent the sources to be reconstructed.
Each source $s_j(t)$ follows a different GP with zero mean, and a distinctive spectral mixture kernel. 
That is, $y_{i} = f(t_i) + \epsilon_i$, where 
$f(t) = \sum_{j=1}^{J} s_{j}(t)$, and
\begin{align}
\label{e.source_function}
s_j(t) &\sim \mathcal{GP}
\left(\
0,
\ k_j(t, t')
\ \right)
\quad  \text{for } j=1,2,\dots, J.
\end{align}
Here, the noise follows $\epsilon_i \sim \mathcal{N}(0, \ \nu^2)$, with variance $\nu^2$. The kernel for the $j$-th source is represented by $k_j(t, t')$ (introduced shortly in section \ref{s.kernel}).
In addition, it is a well known property that the sum of GPs is also a Gaussian process \cite{rasmussen05}. Therefore, the mixture function follows
\begin{equation}
\label{e.mixture_function}
f(t) \sim  \mathcal{GP} 
\left( \
0,  \ \sum_{j=1}^{J}  k_j(t, t') \ 
\right),
\end{equation}
where its kernel is the sum of source kernels, i.e. $k_f(t, t') = \sum_{j=1}^{J}  k_j(t, t') $.
We focus only on predicting the mixture function \eqref{e.mixture_function} as well as the sources \eqref{e.source_function} evaluated at  $\mathbf{t}$.

Any finite set of evaluations of a GP function follows a multivariate normal distribution \cite{rasmussen05}. Therefore, the prior over the mixture function, and each source evaluated at $\mathbf{t}$, correspond to
$\mathbf{f} \sim  \mathcal{N}\left(\mathbf{0}, \ \mathbf{K}_{f} \right)$,
and 
$\mathbf{s}_j \sim \mathcal{N}\left(\mathbf{0}, \ \mathbf{K}_{s_j} \right)$ respectively, 
where the column vectors $\mathbf{f} = [f(t_1), \dots, f(t_n)]^{\top}$,  $\mathbf{s}_j= [s_j(t_1), \dots,  s_j(t_n)]^{\top}$, and the covariance matrix $\mathbf{K}_f = \sum_{j=1}^{J} \mathbf{K}_{s_j}$. 
The matrices $\left\lbrace \mathbf{K}_{s_j}\right\rbrace_{j=1}^{J} $ are computed by evaluating the source kernels at all pair of time instants. That is, $\mathbf{K}_{s_j}[l,l'] = k_j(t_l, t_{l'})$ for $l=1,2,\dots,n$, and $\l'=1,2,\dots,n$.
Also, when a Gaussian likelihood is assumed,
the priors are conjugate to the likelihood \cite{rasmussen05}. Hence, the posterior distributions are also Gaussian. That is,
\begin{align}
\label{e.likelihood}
\mathbf{y} \ | \ \mathbf{f} 
&\sim \prod_{i=1}^{n} \mathcal{N}\left( y_i \ | \ f_i, \ \nu^2\right) ,
\\
\label{e.posterior}
\mathbf{f} \ | \ \mathbf{y} &\sim \mathcal{N} \left(\mathbf{f}_{\text{ }\text{  }} \ | \ \mathbf{K}_{f}^{\top} 
\mathbf{H}^{-1}
\mathbf{y}, \ \ \hat{\mathbf{K}}_f \right),
\\
\label{e.posterior_source}
\mathbf{s}_j \ | \ \mathbf{y} &\sim
\mathcal{N} \left( \mathbf{s}_i \ | \ \mathbf{K}_{s_j}^{\top} 
\mathbf{H}^{-1}
\mathbf{y}, \ \ \hat{\mathbf{K}}_{s_j} \right).
\end{align}
Here, the likelihood \eqref{e.likelihood} factorizes across the mixture data, 
and the posterior over the mixture function \eqref{e.posterior} has covariance matrix 
$\hat{\mathbf{K}}_f= \mathbf{K}_{f} - \mathbf{K}_{f}^{\top} \mathbf{H}^{-1} \mathbf{K}_{f}$.
%
Also, the posterior distribution over the $i$-th source \eqref{e.posterior_source} has covariance matrix $\hat{\mathbf{K}}_{s_j} = \mathbf{K}_{s_j} - \mathbf{K}_{s_j}^{\top} \mathbf{H}^{-1} \mathbf{K}_{s_j}$, where the matrix $\mathbf{H} = \mathbf{K}_{f} + \nu^2\mathbf{I}$, and $\mathbf{I}$ is the identity matrix. 
Further, the model hyperparameters are usually learned by maximizing the log-marginal likelihood 
\begin{equation}
\label{e.marginal}
\log p(\mathbf{y} ) =
-\frac{1}{2}
\left[
\mathbf{y}^{\top}
\mathbf{H}^{-1}
\mathbf{y}
+
\log
\left| 
\mathbf{H}
\right| 
+
n\log 2\pi 
\right],
\end{equation}
where  $\mathbf{H}$ needs to be inverted.

Although the source separation GP model introduced so far is elegant, its application to large audio signals becomes intractable.
This is because the computational complexity of GP inference scales cubically with the number of audio samples.
%
%
%
%
%
Specifically, learning the hyperparameters by maximizing the true marginal likelihood \eqref{e.marginal} is computationally demanding, as it requires the inversion of a $n \times n$ matrix. 
To overcome the limitations imposed by matrix inversion, we instead maximized a variational lower bound of the true marginal likelihood  \eqref{e.marginal} (introduced shortly in section \ref{s.inference}).
In addition, we divided the mixture data into overlapping frames of size $\hat{n} \ll n$.
%
Finally, to reconstruct the sources, we used the hyperparameters learned for each frame to calculate the true posterior distribution over the sources (eq. \eqref{e.posterior_source}).
The rest of this section is structured as follows. Section \ref{s.kernel} introduces the spectral mixture kernel used for each source. 
Then, section \ref{s.inference} presents the lower bound of the true marginal likelihood we maximized for learning the hyperparameters.

\subsection{Spectral mixture kernels for isolated sources}\label{s.kernel}
The kernel  $k_j(t, t')$ in \eqref{e.source_function} determines the properties of each source $s_j(t)$, that is, smoothness, stationarity, and more importantly, its spectrum. 
To model the typical spectral content of each isolated source, we used spectral mixture kernels \cite{Wilson13}. 
These kernels approximate the spectral density of any stationary covariance function using a Gaussian mixture.
Further, Alvarado et al. \cite{Alvarado17} assumed a Lorentzian mixture instead, resulting in the Mat\'{e}rn-$1/2$ spectral mixture (MSM) kernel
\begin{equation}\label{e.kernel}
k_j(\tau) =
\sigma^{2}_{j}  \exp{\left( -\frac{\tau}{\ell_j} \right)} \times  \sum_{d=1}^{D} \alpha^{2}_{jd} \cos(\omega_{jd} \ \tau),
\end{equation}
where $\tau=|t-t'|$, the set of parameters $\left\lbrace \alpha^{2}_{jd}, \ \omega_{jd} \right\rbrace_{d=1}^{D} $ controls the energy distribution throughout 
all the harmonics/partials of the $j$-th source spectrum.
%
In addition, the variance $\sigma^{2}_{j}$ controls the source amplitude, whereas the lengthscale $\ell_j$ determines how fast $s_j(t)$ evolves in time. 
We grouped all the kernel parameters in the set $\boldsymbol{\theta}_{j} = \left\lbrace \sigma^{2}_{j}, \ \ell_j, \ \left\lbrace \alpha^{2}_{jd}, \ \omega_{jd} \right\rbrace_{d=1}^{D} \right\rbrace $.
We fitted a MSM kernel \eqref{e.kernel} to the spectrum of every source. 
For this purpose, we used training data consisting of one audio recording of each isolated source. 
We denoted the training data as $\left\lbrace \mathbf{g}^{(j)}\right\rbrace_{j=1}^{J}$, where  $\mathbf{g}^{(j)} = [g^{(j)}(x_i)]_{i=1}^{\tilde{n}}$ is the training data vector for the $j$-th source, and  $\mathbf{x} = \left[ x_i \right]_{i=1}^{\tilde{n}}$ is the corresponding time vector.
In addition, because only one single realization $\mathbf{g}^{(j)}$ was available for each source in $\left\lbrace s_j(t) \right\rbrace_{j=1}^{J}$, we assumed the sources  to be covariance-ergodic processes with zero mean \cite{Papoulis91, Shanmugan88, Goulard92}. 
Therefore, their covariances $\left\lbrace C_j(\lambda) \right\rbrace_{j=1}^{J}$ were estimated as the time average 
\begin{equation}
\label{e.estimator}
C_j(\hat{\tau}) = \frac{1}{T}\int_{0}^{T}
g^{(j)}(x + \hat{\tau})
\
g^{(j)}(x)
\
\text{d}x.
\end{equation}
%
Here, $T$ denotes the size (in seconds) of the window used to compute the correlation. We used the discrete version of eq. \eqref{e.estimator}.
%
%
Finally, for every source we then minimized the mean square error (MSE) between the covariance estimator \eqref{e.estimator} and the corresponding MSM kernel \eqref{e.kernel}. That is,
\begin{equation}
\label{e.mse_kernel}
L(\boldsymbol{\theta}_j) = 
\frac{1}{N_{c}}
\sum_{i=1}^{N_{c}}
\left[ 
k_j(\hat{\tau}_i)
-
C_j(\hat{\tau}_i)
\right]^2,
\end{equation}
%
where $N_c$ is the number of points where \eqref{e.estimator} was approximated, and $\boldsymbol{\theta}_j$ is the set of kernel parameters in \eqref{e.kernel}. 

\subsection{Preprocessing and inference}
\label{s.inference}
To reduce the computational complexity of learning the hyperparameters by maximizing the true marginal likelihood \eqref{e.marginal}, we divided the mixture data $\left\lbrace t_i, y_i \right\rbrace_{i=1}^{n} $ into $W$ overlapping frames of size $\hat{n} \ll n$. Therefore, the set of frames corresponded to $\left\lbrace  \hat{\mathbf{t}}^{(w)}, \hat{\mathbf{y}}^{(w)}\right\rbrace_{w=1}^{W} $. 
%
%
In addition, for each mixture frame $\hat{\mathbf{y}}^{(w)}$, we instead maximized the lower bound of the true marginal likelihood, proposed by Titsias \cite{Titsias09} for variational sparse GPs. 
This method depends on a smaller set of  \textit{inducing variables} $\mathbf{u} \in \mathbb{R}^m$, where $m \leq \hat{n}$.
The set $\mathbf{u}$ represents the values of the function $f(t)$ (eq. \eqref{e.mixture_function}) evaluated at a set of \textit{inducing points} $\mathbf{z} = \left[ z_i \right]_{i=1}^{m}$. 
Thus, $\mathbf{u} = \left[ f(z_1), \dots, f(z_m) \right]^{\top}$.
The inducing points $\mathbf{z}$ lie on the same domain as $\mathbf{t}$, i.e. time.
Moreover, the inducing points, 
together with the model hyperparameters 
are learned by minimizing the Kullback-Leibler (KL) divergence between the Gaussian approximate distribution $q(\mathbf{u})$, and the true posterior $p(\hat{\mathbf{f}} \ | \ \hat{\mathbf{y}}^{(w)})$.
This approach leads to the following bound
\begin{equation}\label{e.elbo}
\mathcal{L} \overset{\Delta}{=} \log \mathcal{N} 
\left( 
\hat{\mathbf{y}}^{(w)} | \ \mathbf{0},  \  \mathbf{Q}_{\hat{n}\hat{n}} + \nu^{2}\mathbf{I} \right)
- \frac{1}{2\nu^{2}} \text{tr} \left( \mathbf{K}_{\hat{n}\hat{n}} -\mathbf{Q}_{\hat{n}\hat{n}} \right),
\end{equation} 
where the matrix $\mathbf{Q}_{\hat{n}\hat{n}} = \mathbf{K}_{\hat{n} m} \mathbf{K}_{m m}^{-1}\mathbf{K}_{m \hat{n}}$. Here, the cross covariance $\mathbf{K}_{\hat{n} m}[i, j] = k_f(t^{(w)}_i, z_j) $. 
Similarly,  $\mathbf{K}_{m m}[i, j] = k_f(z_i, z_j)$. Where $t_i^{(w)}= \mathbf{t}^{(w)}[i]$.
Recall that $k_f(t, t')$ is the kernel of the mixture function (eq. \eqref{e.mixture_function}).
In brief, the computational complexity of learning hyperparameters in each frame was reduced from $\mathcal{O}(\hat{n}^3)$, to $\mathcal{O}(\hat{n} m^2)$.

\section{Experimental Evaluation}
\begin{figure*}[ht]
	\centering
	\includegraphics[width=0.99\linewidth]{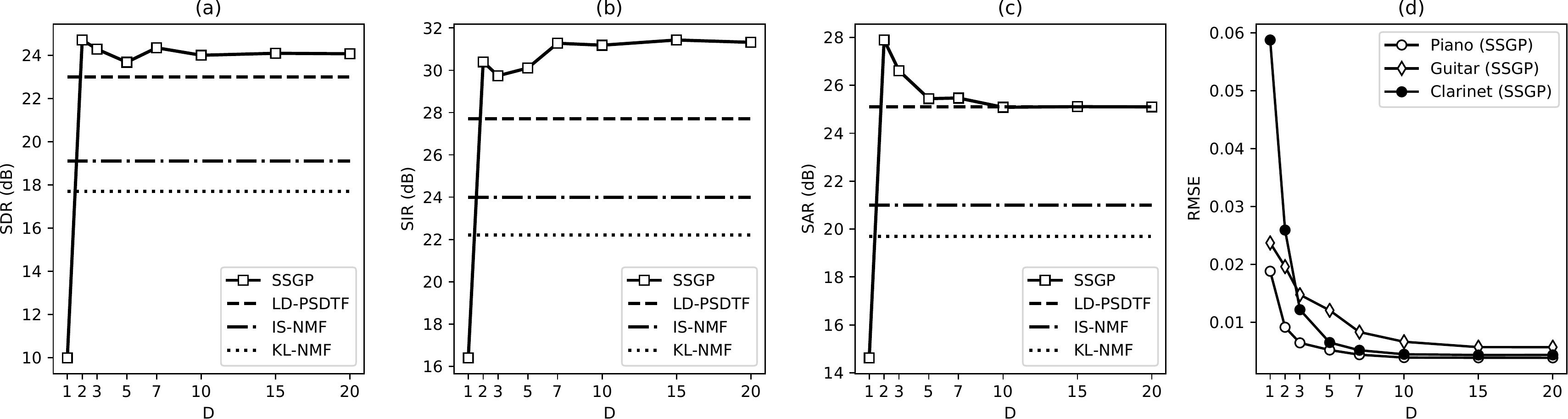}
	\caption{Source separation metrics. SDR (a), SIR (b), SAR (c), RMSE (d).}\vspace{-4pt}
	\label{f.metrics}
\end{figure*}
\begin{table}[t]
	\centering
	\begin{tabularx}{\columnwidth}{lXXXc}
		\hline
		Method   & SDR & SIR  & SAR  &  Opt. time  \\ \hline \hline
		KL-NMF   & 17.7              & 22.2              & 19.7          &  --   \\
		IS-NMF   & 19.1              & 24.0              & 21.0          &  --  \\
		LD-PSDTF & 23.0              & 27.7              & 25.1          &  --  \\ 
		 SSGP \small{(proposed)}    & \textbf{24.1}     & \textbf{31.4}     &
		 \textbf{25.1}      &   \textbf{5.33}      \\ 
		 SSGP-\small{full}    &   22.9     & 22.3     &  
		 24.6      &    284.2    \\ \hline
	\end{tabularx}
    \vspace{0pt}
	\caption{Separation metrics (dB). Optimization time (min).}
	\label{t.eval_metrics}
\end{table}
\begin{figure}[ht]
	\centering
	\includegraphics[width=0.9825\columnwidth]{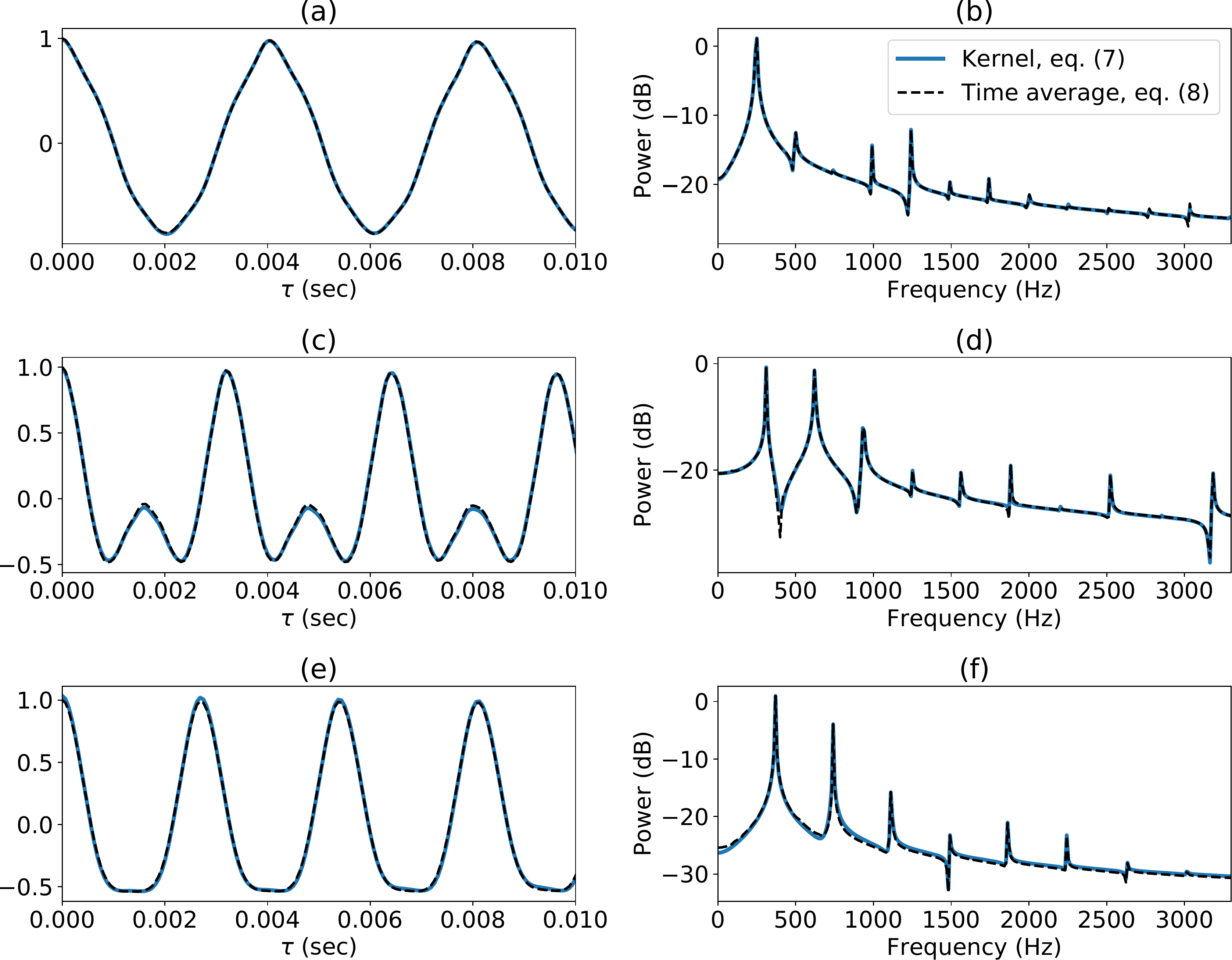}\vspace{-4pt}
	\caption{Kernels learned for piano notes (left column). Corresponding log-spectral density (right column).  }
	\label{f.kernel}
\end{figure}
\begin{figure}[ht]
	\centering
	\subfigure{\includegraphics[width=0.99\columnwidth]{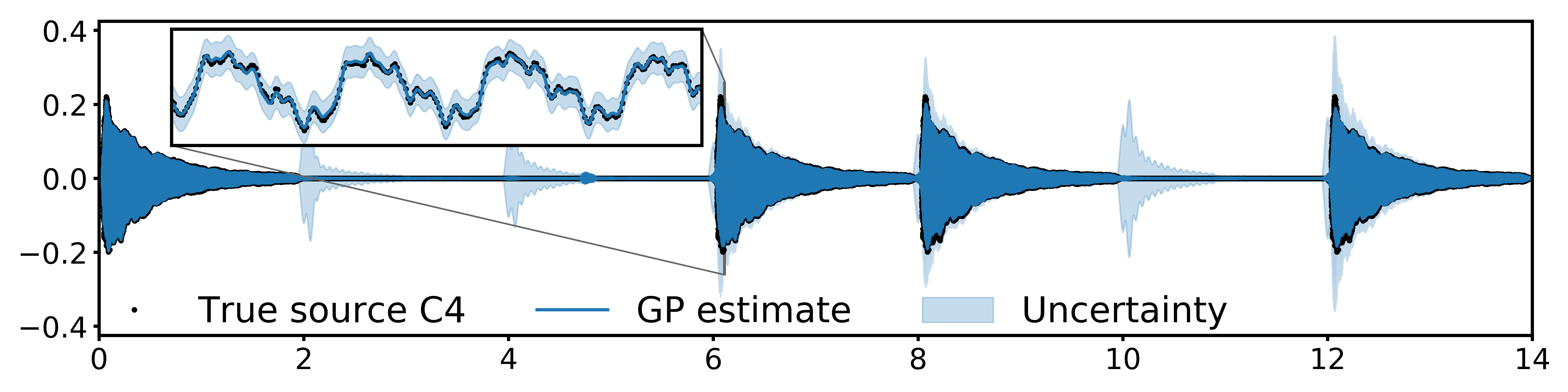}}\vspace{-14.25pt}
	\subfigure{\includegraphics[width=0.99\columnwidth]{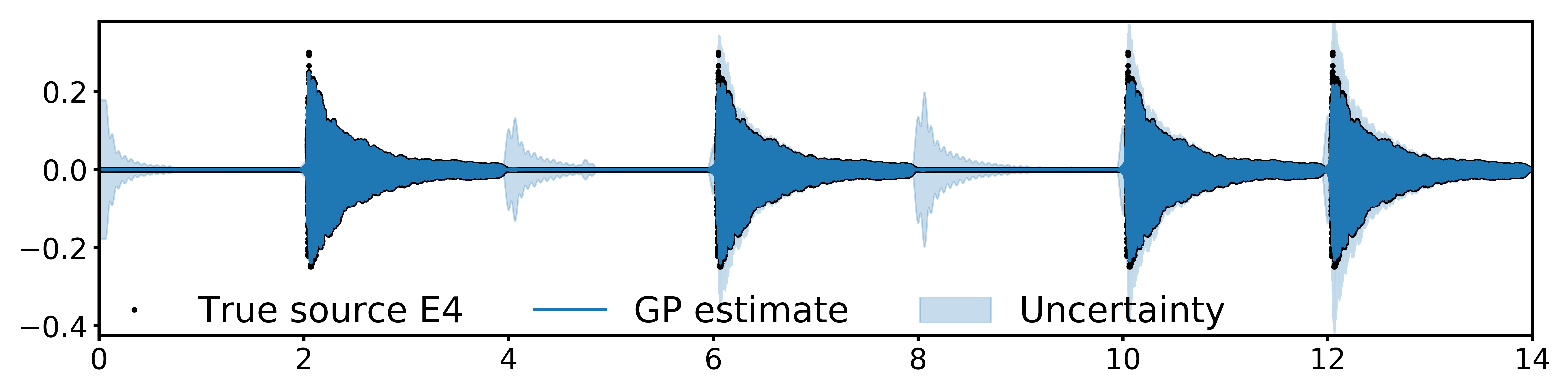}}\vspace{-14.25pt}
	\subfigure{\includegraphics[width=0.99\columnwidth]{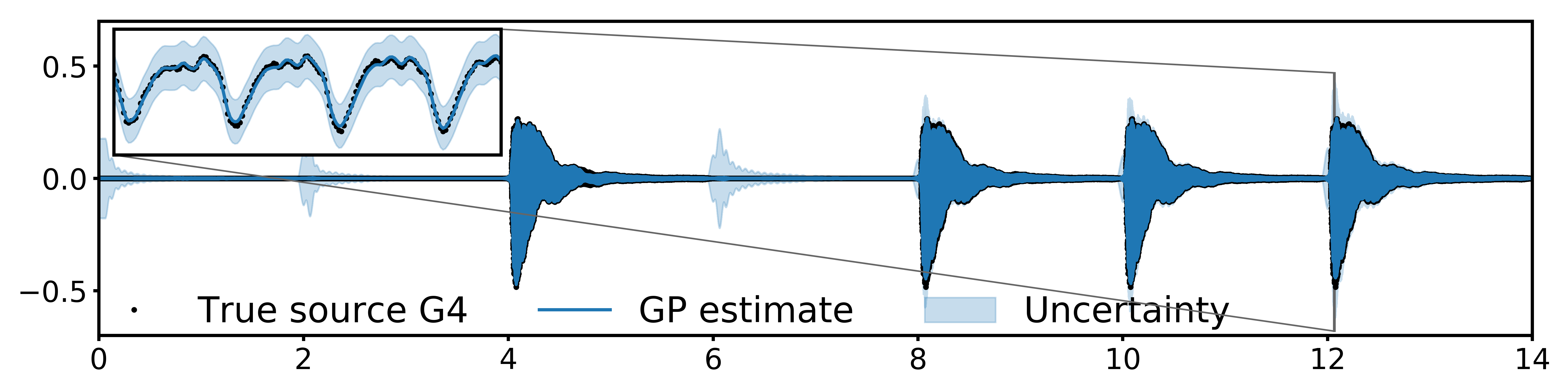}}\vspace{-14.25pt}
	\subfigure{\includegraphics[width=0.99\columnwidth]{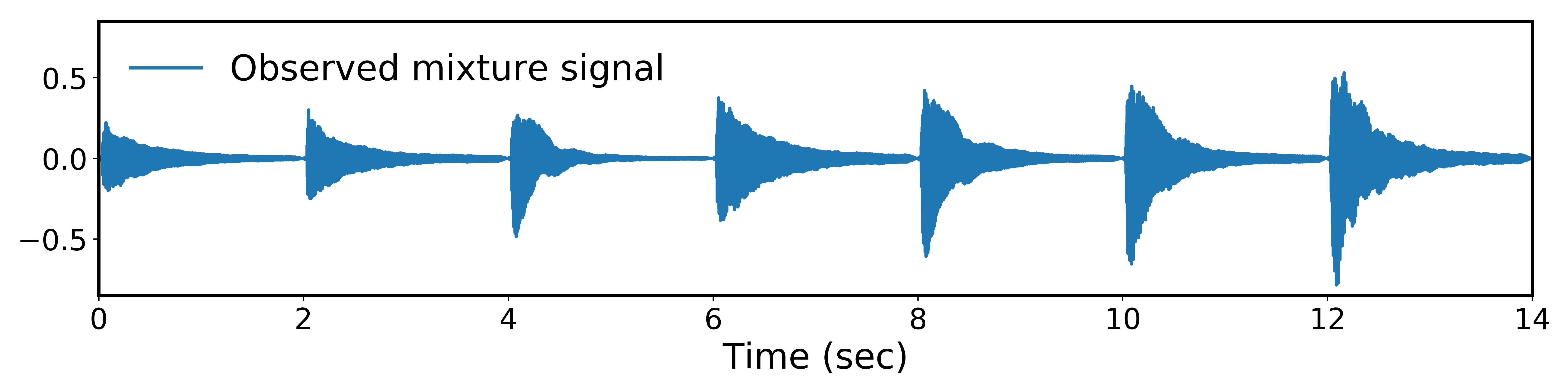}}\vspace{-4pt}
	\caption{Source reconstruction on piano mixture signal.}
	\label{f.separation}
\end{figure}
%
%
We tested the proposed SSGP method on the same dataset analysed in \cite{Yoshii13}.
That is, three different mixture audio signals sampled at $16$KHz, corresponding to piano, electric guitar, and clarinet. 
Each mixture lasts 14 seconds, and consists of the following sequence of music notes (C4, E4, G4, C4+E4, C4+G4, E4+G4, and C4+E4+G4).
Thus, for each mixture, the aim was to reconstruct three source signals, each with a corresponding note, C4, E4, and G4.
The metrics used to measure the separation performance were: source to distortion ratio (SDR), source to interferences ratio (SIR), source to artifacts ratio (SAR) \cite{Vincent06}, and root mean square error (RMSE).
We compared with LD-PSDTF (positive semi-definite tensor factorization),  KL-NMF (Kullback-Leibler NMF), and IS-NMF (Itakura-Saito NMF) \cite{Yoshii13}. 
%
%
The code was implemented using GPflow \cite{Matthews17}. 
%

We determined the performance of the proposed method in mixtures of three sources. That is, $J=3$ in eq. \eqref{e.mixture_function}. 
To this end, we first divided the mixtures into frames of $125$ milliseconds ($\hat{n}=2001$) with $50\%$ overlap, 
and initialized the kernel for each source (eq. \eqref{e.kernel} with $D=15$), by minimizing eq. \eqref{e.mse_kernel}.
Then, for each mixture frame, we maximized eq. \eqref{e.elbo} to learn the variance of each source, i.e., $\left\lbrace \sigma_j^2 \right\rbrace_{j=1}^{J}$.
We used two separate criteria to select $\mathbf{z}$: 
either the inducing points were located at the extrema of the mixture data (sparse GP),
or the inducing points were equal to the time vector (full GP). 
We compared the time required for learning the hyperparameters in these two scenarios.
Finally, we used eq. \eqref{e.posterior_source}, and the learned hyperparameters to calculate the true posterior over each source $p \left(  \mathbf{s}_i^{(w)} | \mathbf{y}^{(w)}  \right)$.
We recovered the sources applying the \textit{overlap}-\textit{add} method to the frame-wise predictions \cite{Allen77}.
%
%
We found that our method (SSGP) presented the highest SDR and SIR metrics, and reduced the optimization time by 98.12\% compared to the full GP (Table \ref{t.eval_metrics}),
indicating that our method is efficient, robust to interferences between sources (highest SIR), and it introduces less distortion (highest SDR).
Further, we observed that the kernels learned for each source presented distinctive spectral patterns (Fig \ref{f.kernel}),
which demonstrates that SM kernels are appropriate for learning the rich frequency content found in audio sources.
Moreover, we observed that the proposed approach reconstructed accurately the sources (Fig \ref{f.separation}),
showing the variances learned by maximizing the lower bound were consistent with the true sources.
%
%
%
%
%
In addition, to establish the effect of kernel selection on the separation performance, 
we carried out the same previous experiment, but changing the number of components $D$ in the kernel eq. \eqref{e.kernel}.
%
We found that SDR, SIR and SAR metrics stabilized when $D > 3$ (Fig. \ref{f.metrics}(a-c)), 
%
indicating that the proposed model is less affected	by kernel selection when more than three components are used. 
Further, RMSE decreased exponentially with $D$ (Fig. \ref{f.metrics}(d)),
suggesting that increasing the number of components in the kernel leads to more accurate waveform reconstructions.

\section{Conclusions}

Our findings indicate that combining variational sparse GPs together with SM kernels enables time-domain source separation GP models to reconstruct audio sources in an efficient and informed manner, without compromising performance.
Also, RMSE results imply that suitable spectrum priors over the sources are essential to improve source reconstruction. 
Moreover, SDR, SIR, and SAR results suggest the proposed method can be used for other applications such as multipitch-detection, where low interference between sources (SIR) is more relevant than reconstruction artifacts (SAR).   
We proposed an alternative method that circumvents phase approximation by addressing audio source separation from a variational time-domain perspective.
The code is available at \cite{Alvarado18}.

\bibliographystyle{IEEEbib}
\bibliography{icassp_bib}

\end{document}